\documentclass[aps,prl,reprint,showpacs,floatfix]{revtex4-1}
\usepackage{graphicx}
\graphicspath{{./3P2Feshbach_Figures/}}
\begin{document}
\newcommand{\ket}[1]
{|#1 \rangle}
\newcommand{\bra}[1]
{\langle #1 |}
\newcommand{\singlet}[0]
{{}^{1}S_{0}}
\newcommand{\triplet}[0]
{{}^{3}P_{2}}
\newcommand{\gup}[0]
{g_{\uparrow}}
\newcommand{\gdown}[0]
{g_{\downarrow}}
\newcommand{\gnd}[0]
{g}
\newcommand{\gupdown}[0]
{g_{\uparrow,\downarrow}}
\newcommand{\exc}[0]
{e}

\title{Feshbach-Resonance-Enhanced Coherent Atom-Molecule Conversion with Ultra-Narrow Photoassociation Resonance}
\author{Shintaro Taie}
\altaffiliation{Electronic address: taie@scphys.kyoto-u.ac.jp}
\affiliation{Department of Physics, Graduate School of Science, Kyoto University, Japan 606-8502}
\author{Shunsuke Watanabe}
\affiliation{Department of Physics, Graduate School of Science, Kyoto University, Japan 606-8502}
\author{Tomohiro Ichinose}
\affiliation{Department of Physics, Graduate School of Science, Kyoto University, Japan 606-8502}
\author{Yoshiro Takahashi}
\affiliation{Department of Physics, Graduate School of Science, Kyoto University, Japan 606-8502}
\date{\today}

\begin{abstract}
We reveal the existence of high-density Feshbach resonances in the collision between the ground and metastable states of
$^{171}$Yb and coherently produce the associated Feshbach molecules by photoassociation.
The extremely small transition rate is overcome by the enhanced Franck-Condon factor of the weakly bound Feshbach molecule,
allowing us to observe Rabi oscillations with long decay time between an atom pair and a molecule in an optical lattice.
We also perform the precision measurement of the binding energies, which characterizes
the observed resonances.
The ultra-narrow photoassociation will be a basis for practical implementation of optical Feshbach resonances.
\end{abstract}
\maketitle
Remarkable advances in study on ultracold atomic gases owe much to their high controllability over
various experimental parameters.
A tunability of the interactions between atoms with Feshbach resonances (FRs) \cite{Chin2010} is undoubtedly
an important key to many experiments. One of the most significant applications of FRs is to create ultracold molecules.
By adiabatically sweeping a magnetic field across a resonance, one can create ultracold, near-threshold
Feshbach molecules, which plays a key role in many studies such as
the crossover between a Bose-Einstein Condensate (BEC) and a Bardeen-Cooper-Schriefer superfluid
\cite{Regal2004,Bartenstein2004,Zwierlein2004},
the initial step to form an ultracold, ro-vibronic ground state of (polar) molecules \cite{Ni2008,Lang2008},
correlation measurements for ultracold atoms in an optical lattice \cite{Joerdens2008,Schneider2008,Greif2011}, and so on.
Photoassociation (PA) is another standard method to generate ultracold molecules which can create
various molecules in electronic excited states with a simple optical excitation. \cite{Jones2006}.
However, the transition probability is never large enough even for a strong electric-dipole allowed (E1) transition
due to small overlap between the wave functions of free atoms and molecular bound states, resulting in small
Franck-Condon factors.
In addition, the radiative lifetime of the created molecular states is usually quite short for an E1 transition.  
As a consequence, coherent production of molecules by one-color PA has been impossible until recent demonstration
with a BEC of $^{88}$Sr \cite{Yan2013}, where the narrow $\singlet \leftrightarrow{}^{3}P_{1}$ transition
was exploited.

In this Letter, we report on photoassociative creation of ultracold molecules associated with the ground $\singlet$ and
long-lived metastable $\triplet$ states of $^{171}$Yb with a radiative lifetime of longer than one second.
Importantly, the otherwise quite small strength of the PA resonance is significantly enhanced
by working around FRs for the closed channel.
This enables us to demonstrate Rabi oscillations in the PA transition with a lifetime reaching hundreds of microseconds.
Moreover, by using this narrow-line PA, observed FRs are characterized by precise measurement of
the binding energies of near-threshold Fesbach molecules.
Our observation of the strong PA line with a long expected radiative lifetime opens up
the possibility to suppress atom loss in optical Feshbach resonances \cite{Theis2004,Enomoto2008b,Yan2013}.

First, we describe our observation of FRs in collisions between $\singlet$ and $\triplet$ states of $^{171}$Yb, 
which is an important basis in this work.
Most of the currently studied FRs are for alkali atoms, which originate from their hyperfine structures
with isotropic van der Waals interactions.
On the other hand, the emergence of FRs associated with the existence of anisotropic interactions was recently predicted
\cite{Petrov2012} and observed in collisions of Er \cite{Frisch2014} and Dy \cite{Baumann2014}.
For these atoms, anisotropy in an electrostatic van der Waals potentials with non-zero electronic orbital angular momenta
as well as dipole-dipole interactions induces couplings between an open channel and many closed channels with
higher partial waves, leading to an extremely high density of FRs \cite{Kotochigova2014}.
In the present case of the collision between the $\singlet$ and $\triplet$ states of Yb atoms,
anisotropy originates purely form the van der Waals interaction with an orbital angular momentum \cite{Kato2013}.

\begin{figure}[bt]
	\includegraphics[width=85mm]{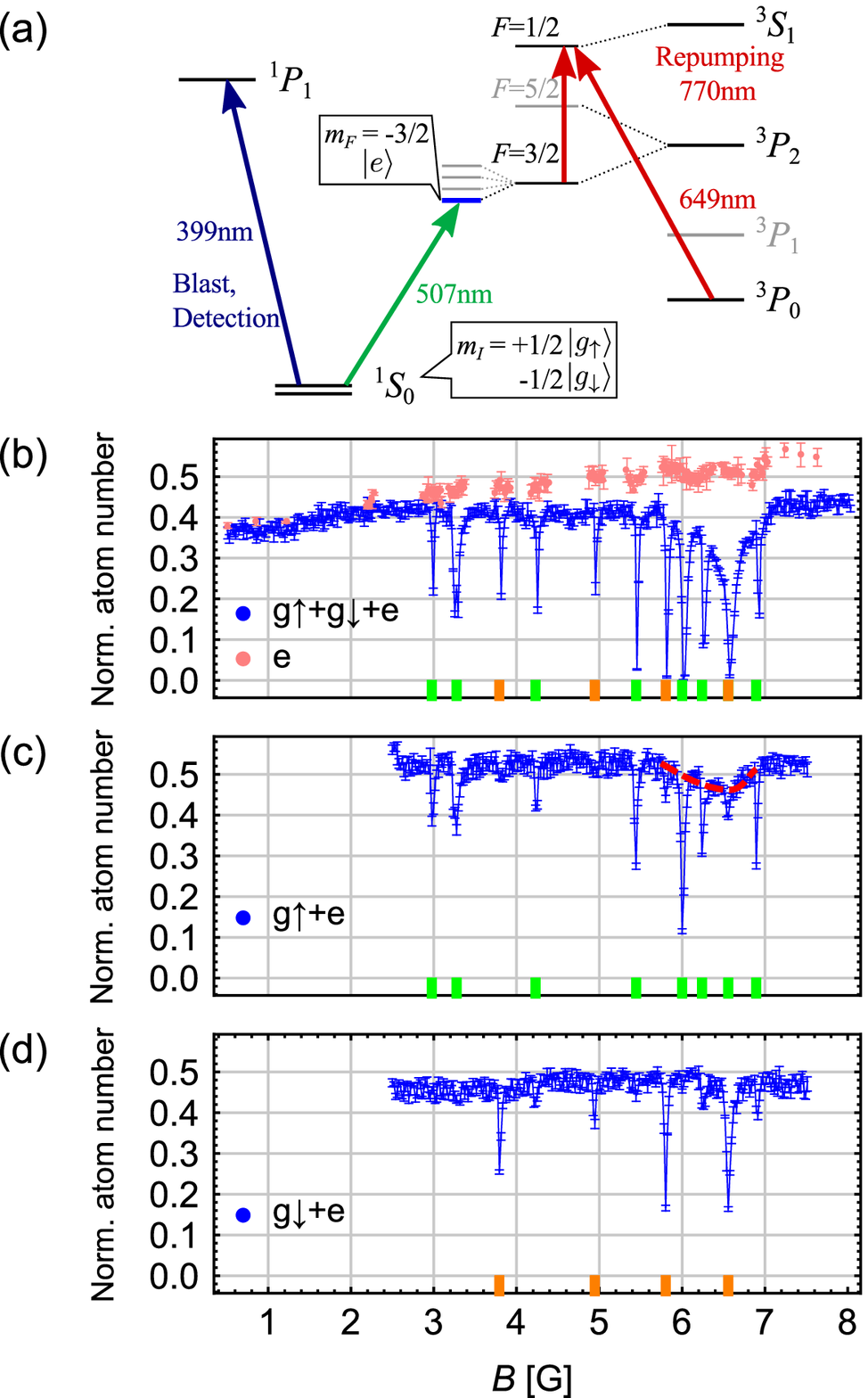}
	\caption{
		(Color online).
		(a) Low-lying energy level diagram of $^{171}$Yb.
		(b-d) Observation of FRs (marked with rectangles) in a mixture of $\gup$-$\gdown$-$\exc$,  $\gup$-$\exc$
		and $\gdown$-$\exc$, respectively.
		Trap loss spectra at temperature of (b) $1.1$~$\mu$K, and (b,c) $1.8$~$\mu$K are shown.
		Pink points in (b) represents the data for a sample containing $\ket{\exc}$ atoms only.
		Red dashed line in (c) highlights the existence of a broad resonance at $6.6$~G. 
	}
	\label{fig_loss}
\end{figure}

We start from the results of trap loss spectroscopy.
Figure \ref{fig_loss} (a) shows the relevant energy diagram and the experimental sequence.
An ultracold Fermi gas of ${}^{171}\text{Yb}$ with two spin components ($\ket{\gup} = \ket{\singlet (m_I=+1/2)}$
and $\ket{\gdown} = \ket{\singlet (m_I=-1/2)}$) is prepared by sympathetic evaporative cooling
with $^{173}\text{Yb}$ in an optical dipole trap with a wavelength of $532$~nm \cite{Taie2010}.
After evaporation we obtain a thermal gas of $8.3(2) \times 10^4$ atoms at a temperature $T = 1.1(1)$~$\mu$K.
The trap frequencies are $( \omega_x, \omega_y, \omega_z )/2\pi = (327, 62.4, 426)$~Hz
with $z$ indicating the direction of gravity, leading to the Fermi temperature of $T_{\rm F} =620$~nK.  
Remaining $^{173}$Yb atoms are removed by $556$~nm light at the resonant frequency
of the $\singlet \rightarrow {}^{3}P_1 (F=7/2)$ transition.
At a static magnetic field of $B=B_z=1.0$~G, we excite a small fraction of atoms to the lowest hyperfine state in $\triplet$,
$\ket{\exc}=\ket{\triplet (F=3/2, m_F=-3/2)}$ state, by applying resonant laser light at $507$~nm \cite{Yamaguchi2010}.
Immediately after the excitation, we ramp up the field to the desired value within $10$~ms.
After a hold time of $100$~ms in the dipole trap, atoms in the ground state is removed,
followed by repumping of survived atoms in the $\triplet$ state into the ground state.
The repumped atoms are detected by fluorescence imaging with a magneto-optical trap
using $399$~nm cooling light.

Figure \ref{fig_loss} (b) shows 12 resonant losses which are attributed to the existence of FRs in
$\ket{\gup}$-$\ket{\exc}$ or $\ket{\gdown}$-$\ket{\exc}$ collisions \cite{PRLsupplement01}.
We exclude the possibility of $\ket{\exc}$-$\ket{\exc}$ resonances by repeating the same experiment with
$\ket{\exc}$ atoms only. Loss spectra for the $\ket{\gup}$-$\ket{\exc}$ and $\ket{\gdown}$-$\ket{\exc}$
states are also shown in Fig. \ref{fig_loss} (c) and (d), respectively, 
by which we can identify the relevant spin combination for an open channel in each resonance.
For spin polarization, we apply optical pumping with the $\singlet \rightarrow {}^1P_1 (F=1/2)$ transition
at the early stage of evaporative cooling.
Due to the incompleteness of optical pumping, resonances with the residual spin component are also visible for each case
of two-component spectrum. We estimate the ratio of atoms in each state to be
$(\gup, \gdown, \exc ) = (30\%, 50\%, 20\%)$, $(70\%, 10\%, 20\%)$ and $(5\%, 75\%, 20\%)$
for Fig. \ref{fig_loss} (b), (c) and (d), respectively, which accounts for the contamination of weak resonances.

We note that the strong loss feature at $6.6$~G seen in Fig. \ref{fig_loss} (b) is the consequence of two overlapping resonances:
the broad resonance in the $\ket{\gup}$-$\ket{\exc}$ collision and the narrow resonance
in the $\ket{\gdown}$-$\ket{\exc}$ collision.
As shown in Fig. \ref{fig_loss} (c), the former FR is accompanied by relatively small loss compared to the other narrow ones,
which possibly reflects the open-channel dominated feature of this resonance. 
The broad resonance with small loss is favorable for practical use.
Exploring FRs for fermionic isotopes is of great importance because the stability of
resonantly interacting two-component Fermi gases provides much wider practical applications.
Especially in the case of $\singlet$-$\triplet$ resonances, using polarized fermions is the best way to avoid inelastic collisions
in the $\triplet$ state which is a dominating decay channel \cite{Uetake2012}.

\begin{figure}[bt]
	\includegraphics[width=85mm]{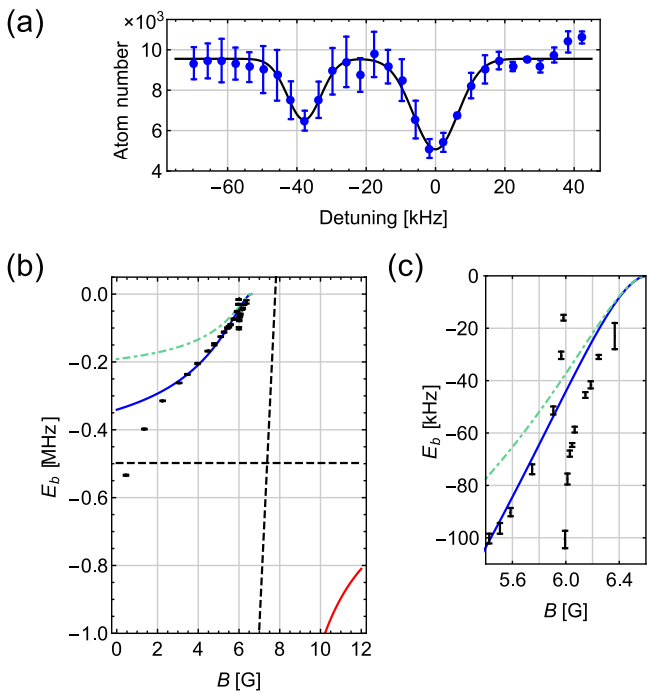}
	\caption{
		(Color online).
		(a) $\singlet \leftrightarrow \triplet$ laser spectroscopy at $B=6.2$~G.
		The PA resonance to the Feshbach molecule associated with the $6.6$~G broadest FR is visible
		at $-38$~kHz from the atomic resonance. The PA laser pulse has a duration $20$~ms and intensity $2.3$~W/cm$^2$.
		(b) Binding energy of the Feshbach molecule associated with resonances at $B=6.0$~G and $6.6$~G.
		The solid lines are the result of the coupled two-channel calculation. Two dashed lines indicate the bare closed
		channel molecule and the shallowest bound state of the open channel. The dot-dashed line is the prediction of the universal relation
		$E_{\rm b} = -\hbar^2/ma^2$. 
		(c) Detailed behavior near the anti level-crossing at $6$~G.
	}
	\label{fig_PA}
\end{figure}

By working around the observed FRs between the ground and metastable states of $^{171}$Yb,
we directly form ultracold molecules with a photo-associative method with significantly enhanced rate.
Owing to the ultranarrow $\singlet \leftrightarrow \triplet$ line in our PA method, it is possible to resolve molecular resonance
down to $\sim 10$~kHz, which is Doppler-limited linewidth of our samples.
Figure \ref{fig_PA} (a) shows a $\ket{\gdown} \rightarrow \ket{\exc}$ laser spectrum taken at $B=6.2$~G, in the vicinity of 
broad $\ket{\gup}$-$\ket{\exc}$ resonance at $6.6$~G. In this measurement, we simply irradiate excitation laser onto
a ground state $\ket{\gup}$-$\ket{\gdown}$ mixture at $T = 65(6)$~nK ($T/T_{\rm F}=0.27$) and measure
the number of atoms remaining in the ground state by absorption imaging.
On the red-detuned side of the atomic resonance, we clearly observe a resonant loss of atoms, which can be attributed to the
association of Feshbach molecules.
Large spatial extent of the Feshbach molecule provides the large Franck-Condon factor, giving PA rate comparable
to the atomic transition. 
Later we return to this point in the context of atom-molecule Rabi oscillations in an optical lattice. 

The frequency detuning of the PA resonance from the atomic resonance directly gives the binding energy
of the associated molecules.
We systematically measure the magnetic field dependence of the binding energy and confirm that the observed molecular state
is associated with the FR at $6.6$~G, as shown in Fig. \ref{fig_PA} (b). 
In addition, we find another molecular level stems from the resonance at $6.0$~G and its anti level-crossing with one from
the $6.6$~G resonance (Fig. \ref{fig_PA} (c)). 
For further characterization of the observed FR, we perform coupled two-channel calculation \cite{Goral2004} of the binding energy.
In the calculation, we fix the unknown parameters $C_6 = 3000$~a.u. for the $\singlet$-$\triplet$ inter-atomic potential and
the differential magnetic moment between the open and the closed channel $\delta \mu = 0.9 \mu_{\rm B}$
where $\mu_{\rm B}$ is the Bohr magneton.
We find that the calculation with background scattering length $a_{\rm bg} = 290 a_0$ and the resonant width $\Delta = 1.11$~G
well reproduce the experimentally determined binding energy. In the low field $B<3$~G, the data shows deviation from the theory,
which indicates the existence of the coupling to unknown molecular levels.
As seen in the figure, the binding energy of the Feshbach molecule remarkably deviates from that of the bare closed-channel molecule.
Therefore the binding energy is not sensitive to the magnitude of $\delta \mu$, which prevents from extracting the
magnetic moment of the closed channel. 

To directly observe the induced change of the scattering length, we perform laser spectroscopy on the
$\singlet \leftrightarrow \triplet$ transition also in an optical lattice. In a deep optical lattice where each lattice site
can be regarded as an isolated potential well, resonance from doubly occupied sites is shifted by the on-site interaction
$U_{\gnd,\exc}-U_{\rm \gnd,\gnd}$ which can be converted to the scattering length $a_{\gnd,\exc}$
\cite{Stoeferle2006,Kato2013}. We adiabatically load a degenerate spin mixture of $\ket{\gup}$ and
$\ket{\gdown}$ at $T/T_{\rm F} = 0.17(2)$ into an simple cubic optical lattice with a wavelength $\lambda=532$~nm.
After reaching the lattice depth of $14.7$ times the recoil energy $E_{\rm R} = \hbar^2 (2\pi/\lambda)^2 /2m$,
we apply an excitation pulse with intensity $0.12 $~W/cm$^2$ and duration $1$~ms to excite $\ket{\gdown}$ atoms,
followed by the repumping scheme to count $\ket{\exc}$ atoms.
Due to the Pauli principle, only the combination of $\ket{\gup}$-$\ket{\exc}$ is produced from doubly occupied sites
and the corresponding scattering length $a_{\gup,\exc}$ can be measured. 

\begin{figure}[bt]
	\includegraphics[width=85mm]{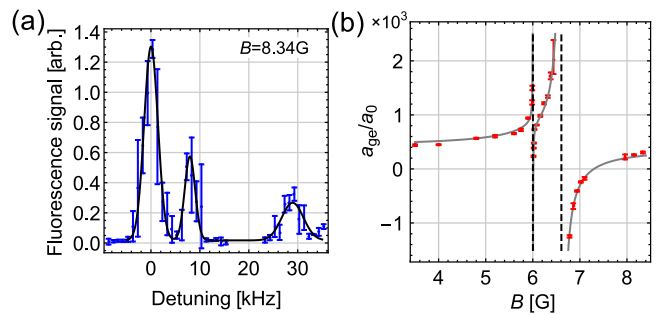}
	\caption{
		(Color online).
		(a) Spectrum on the $\singlet \leftrightarrow \triplet$ transition in an optical lattice with the depth of $14.7 E_{\rm R}$.
		(b) Magnetic field dependence of the scattering length $a_{\rm g\uparrow,e}$ determined by spectroscopy in an optical lattice.
	}
	\label{fig_spec}
\end{figure}

Figure \ref{fig_spec} (a) is a typical spectrum in the lattice, which shows three prominent peaks.
The largest peak corresponds to the excitation of isolated atoms, accompanied by the blue sideband
resonance separated by the band gap $\sim 29$~kHz. 
One more resonance found between these two resonances is the signal from the doubly occupied sites.
We determine the scattering length $a_{\gup, \exc}$ by using the analytic model of a harmonically trapped atom pair
\cite{Busch1998} as
\begin{equation}
a_{\gup,\exc} = \frac{a_{\rm ho}}{\sqrt{2}}
\frac{\Gamma \left( -U_{\gup,\exc}/2\hbar \omega-1/2 \right) }
{\Gamma \left( -U_{\gup,\exc}/2\hbar \omega \right)},
\end{equation}
where the onsite interaction $U_{\gup,\exc}$ is given by the $U_{\gup,\exc} = 2\pi \hbar \Delta f + U_{\gup,\gdown}$ with
$\Delta f$ denoting the frequency difference between the singly and the doubly occupied sites and
$U_{\gup,\gdown} = -2\pi \hbar \times 82$~Hz the onsite interaction in the ground state obtained by the known
scattering length $a_{\gup,\gdown} = -0.15$~nm \cite{Kitagawa2008}.
The mean trap frequency $\omega/2\pi=28.3$~kHz of each potential well is deduced from the band calculation
\cite{PRLsupplement01}.

We map out the magnetic field dependence of $a_{\gup,\exc}$ around the $6.6$~G FR, as shown in Fig. \ref{fig_spec}.
In addition to the resonant change of the scattering length at $6.6$ G, the effect of the narrower $6.0$~G resonance is also visible.
We fit the obtained field dependence with the model proposed in Ref. \cite{Jachymski2013},
\begin{equation}
a(B) = a_{\rm bg} \left( 1-\frac{\Delta_2}{B-B_2} - \frac{\alpha \Delta_1}{B-B_1} \right) \label{eq_amodel}
\end{equation}
where $\alpha = [ (B^{(0)}_1-B^{(0)}_2)/(B^{(0)}_1-B^{(0)}_2-\delta B_2) ]^2$.
Here $B^{(0)}_1(\simeq B_1)$ and $B^{(0)}_2$ are the zero-crossing points of bare molecular energies.
For the shift $\delta B_2=B_2-B^{(0)}_2$, we adopt an approximation \cite{Goral2004}
\begin{equation}
\delta B_2 \simeq \Delta_2 \frac{(a_{\rm bg}/\bar{a})(1-a_{\rm bg}/\bar{a})}{1+(1-a_{\rm bg}/\bar{a})^2} 	
\end{equation}
with the mean scattering length $\bar{a} = 4.4$~nm evaluated from the $C_6$ constant mentioned above.
Equation \ref{eq_amodel} is valid when the width of the first resonance is much narrower than that of the second resonance,
$|\Delta_1| \ll |\Delta_2|$. The fit yields $B_1 = 6.003(5)$~G, $\Delta_1 =3.7(3)$~mG,
$B_2 = 6.607(9)$=G, $\Delta_2 = 680(40)$~mG, and $a_{\rm bg} = 406(15)a_0$. 
Obtained parameters slightly differ from that inferred from the binding energy measurement.
Possible source of the uncertainty in the scattering parameters is the existence of unknown resonances at higher fields
\cite{PRLsupplement01}.
Especially, extracting the value of $a_{\rm bg}$ from the fit is highly sensitive to tails of unknown resonances.

Narrow linewidth of the PA transition and resulting long lifetime of an associated molecule enables coherent
production of molecules.
For the narrow line ($\Gamma /2\pi = 7$~kHz) PA of a  ${}^{88}$Sr BEC, coherent
Rabi oscillations up to $\sim 10$~$\mu$s were reported \cite{Yan2013}. 
Isolated atom pairs residing on each site of an optical lattice are more suitable system to observe coherent phenomena,
for unwanted inelastic collisions are suppressed and long lifetime of molecules can be achieved \cite{Chotia2012}.

\begin{figure}[bt]
\includegraphics[width=85mm]{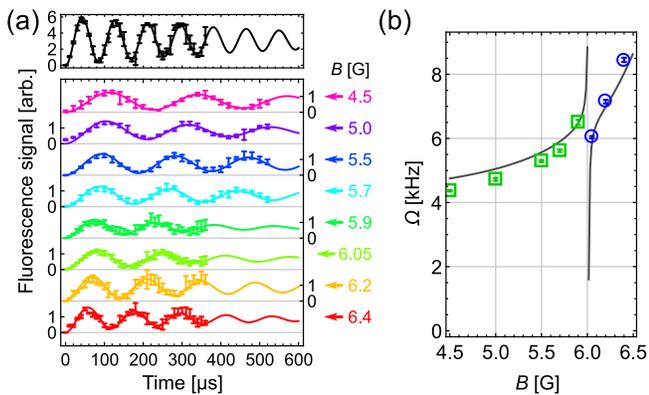}
\caption{
(Color online).
(a) Rabi oscillations of isolated atoms (top) and atom-pairs $\leftrightarrow$ molecules at various magnetic field (bottom) in an optical lattice. The laser intensity is fixed to $58$~W/cm$^2$.
For the atomic transition, the Rabi frequency is $11.8$ kHz.
(b) The atom-molecule Rabi frequency $\Omega$ as a function of the magnetic field. Symbols (circles, squares) distinguish the
different branches of the anti-crossing at $6$~G. The solid line shows the prediction from the wave function overlap (see text).
}
\label{fig_Rabi}
\end{figure}

We demonstrate the Rabi oscillations between $\ket{\gup}$-$\ket{\gdown}$ atom pairs and $\ket{\gup}$-$\ket{\exc}$
Feshbach molecules via one-color PA.
Figure \ref{fig_Rabi} (a) shows both the atomic and atom-molecule Rabi oscillations in the optical lattice
with $14.7 E_{\rm R}$ depth.
For the $\ket{\gdown} \rightarrow \ket{\exc}$ transition is E1 allowed, the Rabi frequency is $\sim 10$ times larger
than the purely M2 allowed case \cite{Porsev2004b}, which makes it much easier to observe clear oscillations.
The $e^{-1}$ decay time of the oscillations is on the order of $500$~$\mu$s, which is limited by the inhomogeneous broadening
of the resonant frequency.
The observed atom-molecule Rabi frequency is on the same order as that of the atomic transition
and keeps reasonable magnitude over the range of several gauss.
To account for the field dependence of the Rabi frequency, we carry out the following simple analysis.
The optical coupling between the initial and the final state is proportional to $\bra{\gup,\exc} d(R) \ket{\gup,\gdown}$,
where $d(R)$ is the transition dipole depending on the inter-atomic distance $R$ \cite{Napolitano1994}.
For a weakly bound Feshbach molecule, $d(R)$ is almost unchanged from its asymptotic value, {\it i.e.}, the atomic transition
dipole, over the most range of molecular extent. Therefore the Rabi frequency can be reduced to the simple product
$\Omega = \Omega_{\rm atom} \times |\langle \gup,\exc | \gup,\gdown \rangle|$.
Using the analytic expression for the two-particle wave functions in a harmonic potential \cite{Busch1998} and
the experimentally determined scattering length, we find good agreement between the data and the above estimation
(Fig. \ref{fig_Rabi} (b)).

In conclusion, we observed the FRs between the ground and metastable excited states
of $^{171}$Yb and successfully formed associated Feshbach molecules by one-color PA.
Observed density of resonances, $0.7$~/G per spin combination, is lower than the cases of highly magnetic Er and Dy,
while much higher than the typical of alkali atoms.
This may be the consequence of the simpler level structure and less anisotropic interactions of $^{171}$Yb. 
Moderately dense resonances are convenient for tuning interactions with a selected well-behaved resonance and for future
theoretical analysis of the observed resonances.
The Feshbach molecule can be strongly coupled to the ground-state scattering wave function via the PA transition.
Optical Feshbach resonance on this transition will effectively change the interaction between the ground state
with a small atomic loss.
Finally, we note that ultra-narrow PA demonstrated here is applicable not only to the homonuclear collisions of
alkaline-earth-like atoms, but also to the heteronuclear collisions between alkali atoms
and alkaline-earth-like atoms, such as Li-Yb \cite{Hara2011}.

\begin{acknowledgments}
This work was supported by the Grant-in-Aid for Scientific Research of JSPS (No. 25220711, No. 26247064)
and the Impulsing Paradigm Change through Disruptive Technologies (ImPACT) program.
\end{acknowledgments}

\newpage

\section*{Supplemental Material}
In this material, we use the following notation defined in the main paper: 
\begin{eqnarray}
&\ket{\gup} = \ket{\singlet, m_I = +1/2},\\
&\ket{\gdown} = \ket{\singlet, m_I = -1/2},\\
&\ket{\exc} = \ket{\triplet, F = 3/2, m_F=-3/2}.
\end{eqnarray}
\section{Excitation to the $^3P_2$ state}
\begin{figure}[b]
	\includegraphics[width=85mm]{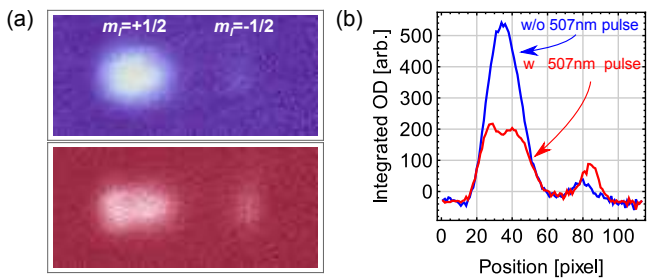}
	\caption{
		(a) Optical Stern-Gerlach experiment for $^{171}$Yb polarized in the $\ket{\gup}$ state by optical pumping before applying
		excitation laser (top) and immediately after excitation to the $\ket{\exc}$ state (bottom).
		Population in the $\ket{\gdown}$ state increases by the stimulated emission from the excited state.
		(b) Atom densities integrated along the vertical direction.}
	\label{fig_OSG}
\end{figure}

While both the $\ket{\gup} \rightarrow \ket{\exc}$ and the $\ket{\gdown} \rightarrow \ket{\exc}$ transitions are allowed
by the magnetic quadrupole (M2) selection rule, the latter transition is much stronger because the hypefine-interaction induced
electric dipole (E1) transition is allowed \cite{Porsev2004bs}. In the low magnetic field, these transitions overlap each other
and we chose the polarization of the excitation laser to suppress the unwanted transition.
However, when we excite $\ket{\gup}$ atoms to the $\ket{\exc}$ state, the imperfection of the polarization
creates a small number of $\ket{\gdown}$ atoms by stimulated emission from the $\ket{\exc}$ state (see Fig. \ref{fig_OSG}).
In the experiment with polarized $\ket{\gup}$ atoms (Fig. 1 (b) in the main paper), we apply the second pumping pulse
after excitation to the $\ket{\exc}$ state, to remove $\ket{\gdown}$ atoms.
This operation heats up atoms by $\sim 500$~nK and is applicable only to non-degenerate thermal atoms.

\section{Loss spectrum at higher magnetic field}
\begin{figure}[bt]
	\includegraphics[width=75mm]{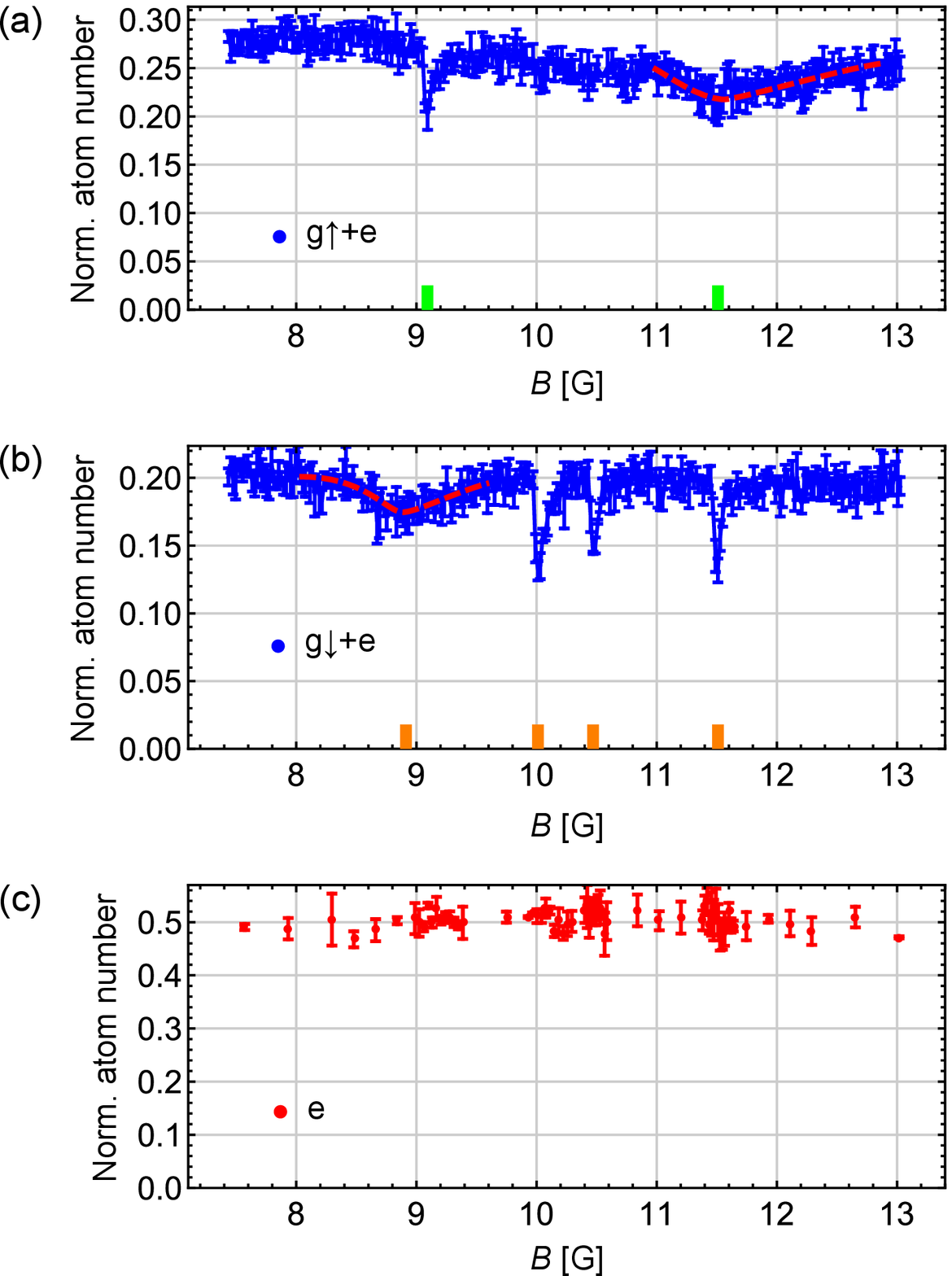}
	\caption{
		Trap loss spectrum for two-component mixtures of (a) $\ket{\gup}$-$\ket{\exc}$ and (b) $\ket{\gdown}$-$\ket{\exc}$.
		(c) The same experiment with purely $\ket{\exc}$ atoms shows the absence of FRs within the $\ket{\exc}$ state.
	}
	\label{fig_loss2}
\end{figure}

In this section, we present the result of atomic loss spectroscopy of Feshbach resonances taken in an extended field range.
As in the case of Fig.~1 (b)-(d) in the main paper, we excite atoms to the $\ket{\exc}$ state, followed by magnetic-field
sweep to variable final values at which atom loss is measured.
For a technical reason, field ramp time after the state preparation is limited to $30$~ms in this range, which causes
additional loss during the sweep. Figure \ref{fig_loss2} shows the 2 and 4 resonances for the $\ket{\gup}$-$\ket{\exc}$
and $\ket{\gdown}$-$\ket{\exc}$ mixtures, respectively. The resonance at $11.5$~G for the $\ket{\gup}$-$\ket{\exc}$
mixture shows broad loss which is comparable to the $6.6$~G resonance studied in the main paper.
Similarly, a broad resonance is also found for the $\ket{\gdown}$-$\ket{\exc}$ mixture at $8.9$~G.

\section{Light shift of the $^3P_2$ state}
Our $532$~nm dipole trap and optical lattices provide the $\ket{\exc}$ state with slightly deeper potential
than the $\ket{\gnd}$ state. While the ac polarizability in the $\ket{\gnd}$ state is independent
of the polarization of the far off-resonant laser light, the light shift of the $\ket{\exc}$ state depends on the angle
$\theta$ between the polarization and the quantization axis \cite{Ido2003,Yamaguchi2010s}.
For linearly polarized laser light, the ac polarizability of a magnetic sublevel $m_F$ in a hyperfine manifold $F$
is given in terms of a scalar and a tensor polarizabilities as
\begin{equation}
\alpha^{(F,I)}_{m_F} =  \alpha_{\rm s}^{(F,I)} + \left( 3 \cos^2 \theta -1 \right) \alpha_{\rm t}^{(F,I)} \frac{3m_F^2-F(F+1)}{2F(2F-1)},
\end{equation}
as long as the light shift can be treated as a small perturbation to the Zeeman splittings \cite{Kien2013}.

\begin{figure}[bt]
	\includegraphics[width=75mm]{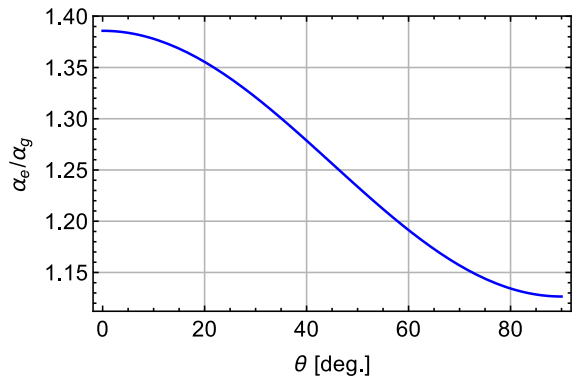}
	\caption{
		The ratio of the light shift in the $\ket{\exc}$ state to that in the $\ket{\gnd}$ state at $532$~nm,
		as a function of the angle between the light polarization and the magnetic field. Here, we assume
		$\alpha_{e}^{\rm (s)}/\alpha_g=1.21$ and $\alpha_{e}^{\rm (t)}/\alpha_g=0.17$
		for the scalar and tensor polarizability in the $\ket{\exc}$ state, respectively.
	}
	\label{fig_lightshift}
\end{figure}

Figure \ref{fig_lightshift} shows the calculated ratio of the light shift in $\ket{\exc}$ state
to that in the $\ket{\gnd}$ state. In our setup, only a lattice beam along the $x$-axis has the polarization
parallel to the magnetic field ($z$), giving rise to the maximum light shift for the $\triplet$ state.
Polarization of all other trapping beams, including the lattice beam along excitation laser ($y$),
point to directions perpendicular to the magnetic field so that the differential light shift
between the states $\ket{g}$ and $\ket{e}$ can be minimized.

In order to determine the scattering length from spectroscopy in an optical lattices (Fig. 3 of the main paper),
we take account of the above light shift and the trap frequency $\omega$ of each potential well is calculated as
$\sqrt{(\omega_{\gnd}^2 + \omega_{\exc}^2)/2}$,
where $\omega_{\gnd} = 2\pi \times 26.6$~kHz for the ground state is obtained by the band calculation and
$\omega_{\exc} = 2\pi \times 29.9$~kHz is the geometric average of the slightly anisotropic trap frequencies
for the $\ket{\exc}$ state.
Broadening ($\sim$ few kHz) of the linewidth of each peak is also caused by the differential light shift.
In this respect, the instability of the magnetic field strength ($\sim 0.5$~mG), and the linewidth of the excitation laser
($< 300$~Hz) also contribute to the measured linewidth.

\end{document}